\documentclass[conference]{IEEEtran}
\usepackage[colorlinks,citecolor=black]{hyperref}     
\hypersetup{citecolor=blue}					
\ifCLASSINFOpdf
\else
\fi
\usepackage{url}
\usepackage[pdftex]{graphicx}
\begin{document}
\title{Optimizing Key Distribution in Peer to Peer Network Using B-Trees }
\author{\IEEEauthorblockN{Abdulrahman Aldhaheri}
\IEEEauthorblockA{School of Engineering and Technology\\Computer Science and Engineering\\
University of Bridgeport\\
Email: \href{mailto:aaldhahe@my.bridgeport.edu}{aaldhahe@my.bridgeport.edu} }
\and
\IEEEauthorblockN{Hammoud Alshammari}
\IEEEauthorblockA{School of Engineering and Technology\\Computer Science and Engineering\\
University of Bridgeport\\
Email: \href {mailto:halshamm@my.bridgeport.edu}{halshamm@my.bridgeport.edu}}
\and
\IEEEauthorblockN{Majid Alshammari}
\IEEEauthorblockA{School of Engineering and Technology\\Computer Science and Engineering\\
University of Bridgeport\\
Email: \href {mailto:maalsham@my.bridgeport.edu}{maalsham@my.bridgeport.edu}}
}
\maketitle


\begin{abstract}
Peer to peer network architecture introduces many desired features including self-scalability that led to achieving higher efficiency rate than the traditional server-client architecture. This was contributed to the highly distributed architecture of peer to peer network. Meanwhile, the lack of a centralized control unit in peer to peer network introduces some challenge. One of these challenges is key distribution and management in such an architecture. This research will explore the possibility of developing a novel scheme for distributing and managing keys in peer to peer network architecture efficiently.
\end{abstract}

\IEEEpeerreviewmaketitle


\hfill
\section{Introduction}
Peer to peer network architecture allows peers to share available resources with each other in a decentralized way \cite{Abraham2005Skip}. It's done efficiently using IP multicasting, which raises concerns about the security of system \cite{Rafaeli2003A-survey}. To provide security to the system, data transmitted has to be encrypted using a key that is known only to peers authorized to access the information. This motivated researchers to find the most efficient way to distribute those keys in order to improve the overall efficiency of the peer to peer system.

On the other hand, B-tree is a very fast and efficient  data structure that is used to store and search large block of data is a logarithmic time. It achieves this by maintaing its balance, and avoiding have great hight. The worst case hight is:

\begin{equation}
	h\quad \le \quad \left\lfloor \log _{ d }{ \quad (\frac { n\quad +\quad 1 }{ 2 } ) }  \right\rfloor
\end{equation}

Where, h is the B-tree hight, d is the maximum number of children a node could have, and n is the number of nodes. This, in fact, provides a feature that could be of great benefit to peer to peer. Having a shallow and balanced tree hierarchy could improve the efficiency of the key management and distribution in peer to peer network. 

\hfill

Because of some characteristics i.e. the small average of failures and laking in central controlling, Peer to Peer (P2P) has been become most popular during these days. However, since there is no such a centralized system is implemented, some of security concerns have beed raised. Decentralized systems, like P2P, have no single server to control the system and play the main role in the whole system. So, by missing that, P2P became applications have beed changed from using simple data to more sensitive data to security threats \cite{Arnedo-Moreno2011Secure}. 

\hfill

Another important aspect is the duration of time that the peer should wait to get the data from the root \cite{Ismail2010A-decision}. In addition, for security purposes that time should not be long and the last nodes should get the session key as fast as the above nodes or so. 

\hfill

This paper proposes designing a B-tree based key distribution and management scheme for peer to peer networks. It will provide higher efficiency rate given the characteristics of B-tree data structure.

\hfill


\section{Related work}

\subsubsection{EKMD}
\hfill

A research group, Liu, et. al. proposed a key distribution and management scheme in peer to peer live streaming network \cite{Liu2007Efficient}. The major properties of given scheme are media-dependent and time-event-driven that the session keys are generated periodically and the re-keying messages are distributed with the media transmission track. The analysis and simulation results demonstrate its properties of security, scalability, reliability and efficiency. It achieves a high performance in security guarantee in p2p live media streaming applications, for which it is very suitable.

\hfill

 An interesting proposal that \cite{Liu2007Efficient} had proposed an efficient media-dependent and time-event-driven key management and distribution scheme, named 'EKMD' for Peer-to-Peer (P2P) live streaming system.
EKMD is Hierarchy Tree Scheme (HTS), centralized approach. It means the SK should be changed once a user joins or leaves the group.
KDC only has to deliver a new SK securely to a small number of group users, which are its immediate neighbors. These neighbors forward the new SK securely to their own neighbor users.

\hfill 

The particular properties of the scheme include: 
\begin{enumerate}
\item Media-Dependent: The key updating (re-keying) messages are embedded into the media content and then delivered through the data transmission track in p2p streaming applications.
\item Time-Event-Driven: The session key updating process is carried out by the Key Distribution Center (KDC) periodically and irrespective of clients' join or leave behaviors. 
\end{enumerate}

\hfill

One of the most important and challenging tasks in peer to peer network is maintaining consistent architecture as users join an leave the network:
\begin{itemize}
\item User Joins:
When a user wants to join the p2p media streaming group, it should firstly contact the KDC to be authenticated. Then it can find its trust neighbors in the group, and get the future re-keying messages from them
\item User Leaves
When a user is going to leave the group, it should firstly notify all its neighbors. After that, the user contacts the KDC to logout.
\end{itemize}

\hfill

Another research group has studied similarity formation of groups and key management in dynamic peer to peer e-commerce \cite{Musau2010Similarity}. This research gave a clear outline of how peers form groups, select group leaders. Then, it addresses the key distribution among groups after selecting a leader for each group. This research also addresses how to establish trusts in peer to peer network.

\hfill

\cite{Liu2005Secure} has presented a simple key distribution protocol, called VTKD (virtual token based key distribution) which was especially designed for collaborative applications to support closed, small dynamic peer group meetings. 
VTKD is a decentralized group key distribution protocol that is based on the Diffie-Hellman (DH) key exchange principle. There is no central group key authority. In contrast to the key exchange between two partners, in the distributed approach each group member calculates a secrete key with each partner using the Diffie-Hellmann principle. 
VTKD is a token based protocol. The group key is renewed whenever the group composition changes (join, leave, and failure of peers)

\hfill

Another important issue we had to review the literature for is the performance of the join operation. The efficiency of the join operation can be measured by the join latency, which is defined by the time difference between the joining peer sending the join request to the server and the joining peer being inserted into the system. To give a quantitative analysis on the join latency, \cite{Yang2010An-Efficient} use the number of hops the join request passes to estimate the join latency.

\hfill

In general, there are two types of peer-to-peer network topologies; structured, and unstructured. \cite{Yang2010An-Efficient} proposed a new approach to get a hybrid topology. The objective of this work is to design a hybrid peer-to-peer system for distributed data sharing which combines the advantages of both types of peer-to-peer networks and minimizes their disadvantages. In their article,  \cite{Yang2010An-Efficient} discusses two main things: 
\begin{enumerate}
\item Leaving and joining the nodes, which we are interesting in.
\item Distributed and sharing data.
\end{enumerate} 

\hfill

The authors separate the nodes into two main categories which are core and not core. The top nodes are connected in structured ring network. The bottom nodes are connected in nun-structured scheme. The core transit network, called t-network, is a structured peer-to-peer network which organizes peers into a ring similar to a chord ring. 
The basic idea behind the hybrid peer-to-peer system proposed by \cite{Yang2010An-Efficient} is that the t-network is used to provide efficient and accurate service while the s-network is used to provide approximate best-effort service to accommodate flexibility.

\hfill

In their research, \cite{Bianchi2010Stabilizing} describe a height balanced tree structure which is Dissemination R-Tree. Each leaf node in the tree is an array of pointers to spatial objects. The joining and leaving nodes relays on some algorithms that's make the tree balanced. The hierarchy is getting changed by apply some strategies like correction of the cover, correction of the level and correction of the tree balance.

\hfill

The research \cite{Bianchi2010Stabilizing} also describe a height balanced tree structure which is Dissemination R-Tree. Each of the node can be the first node, so it can dynamically select the first node to eliminate the case of the first node's failure in a binary tree. We would like to highlight the following points in \cite{Bianchi2010Stabilizing} proposal: 
\begin{itemize}
\item DR-trees generalize P-trees which are the dynamic version of B+ trees.
\item One of the future works that the authors were mentioned about is the time of node of the online:
Loading capacity of nodes are influenced by online time. The model in order to consider it convenient setting the time-line of each node is a constant value.
	
\end{itemize}

\hfill

Using balanced trees to optimize peer-to-peer network has a lot of benefits. A research group proposed a scheme called Skip B-Tree that implement a new algorithm to optimize the load balancing of the files among peers \cite{Abraham2005Skip}. This research proposes a new implementation for a novel data structure called skip b-tree, which is a combination of skip graph and b-tree. The research propose implementing the skip b-tree data structure in peer-to-peer network. The proposed solution would increase the speed and the efficiency of the network.

\hfill

The research proposed by  \cite{Abraham2005Skip} suggest implementing the skip b-tree in allocating resources to peers. However, it doesn't propose implementing the proposed data structure in distributing keys among peers.

\hfill

According to \cite{Graefe2011Modern} , the core design of B-trees has remained unchanged in 40 years: balanced trees, pages or other units of I/O as nodes, efficient root-to-leaf search, splitting and merging nodes, etc. On the other hand, an enormous amount of research and development has improved every aspect of B- trees including data contents such as multi-dimensional data, access algorithms such as multi-dimensional queries, data organization within each node such as compression and cache optimization, concurrency control such as separation of latching and locking, recovery such as multi-level recovery, etc.

\hfill

As suggested by \cite{Beg2008An-Approach,Graefe2011Modern,Beg2009Reduction,Abraham2005Skip,Pang2004Steganographic,Bayer1972Symmetric} The idea of optimizing the original design of B-tree for a specific purpose is not only a valid approach, but also an encouraged one. This actually support our approach in customizing the original B-tree data structure to make it suitable for distributing keys in peer-to-peer network architecture.

\hfill


\section{JOINING AND LEAVING NODES in B-TREE}
Joining and leaving peers in P2P usually happens by following some steps which have been explained in Kwon2007\cite{Kwon2007Public}. These steps illustrated there is no specific rules that control the joining node to determine the parent node based on balancing interesting. Consequently, the main goal of our scope is not presented her which is delivering the session key to the all nodes at the same time or so. 

\hfill

\subsection{JOINING THE NETWORK}
In B-tree architecture, joining nodes happens during two main steps which are searching about the value and split the child \cite{Beg2008An-Approach}. Although, these steps need more work to come up with a balanced B-tree, this additional work still important in terms of security. When node joins, it has to have the permission of joining a group and it has to go in place where keeps the tree balanced either it is leaf or not. Splitting the child means more expanded vertically which gives less number of raws which means that root node will be close to that nodes.

\hfill

\subsection{LEAVING THE NETWORK}
Since there is different ways to implement B-tree, joining and leaving nodes goes through some steps in different strategies. One of these strategies is illustrated in Chang2009A \cite{Chang2009A-Fast} which implies that this node has to go in revers steps on joining nodes. Leaf node doesn't have to be prepared to any situation whereas the upper level nodes have to be prepared to rebuild the tree again.
\hfill


\hfill
\section{Problem statement}
Peer to peer network architecture increases networks efficiency and minimizes bandwidth consumption because it offers a highly decentralized architecture. This high level of decentralization in peer to peer networks increased it’s complexity and imposed security threats on peer to peer network architecture. Data encrypting techniques are implemented to provide security by increasing confidentiality, thus, eliminating the security threats. Encryption and decryption algorithms require having secret key shared between the sender and the receiver. Yet, those keys have to be send encrypted. These tasks are managed by a key distribution. The complexity of the scheme the key distribution center apply affect the efficiency of the peer to peer network significantly. Which would in turn affect the over all performance of the network by consuming more bandwidth.

\hfill

We hypothesize that a B-tree based key distribution scheme can provide better performance in key distribution, which in turn, can lead to more efficient network services. We propose a B-tree based key distribution scheme. We intend to design and implement a key distribution and management scheme for peer to peer network based on B-Trees data structure. We will propose a novel version of B-tree algorithm that is customized to provide faster access time in peer to peer network architecture. We will perform experiments on the proposed scheme by simulating traffic in a control environment. 

\hfill


\section{Expected Outcomes}
\hfill

We are expecting by the end of this project to design and develop a novel key disruption and management scheme that would increase the efficiency of peer to peer network

\hfill

The proposed scheme is also expected to implement a key distribution distribution algorithm, which will be developed using the blueprint of B-tree data structure.

\hfill

The key distribution algorithm is expected to maintain the basic characteristics and functionalities of the B-tree data structure. However, it will be modified and customized to better serve it's purpose within the scope of this research.

\hfill

The newly developed key distribution and management scheme will, then, be evaluated either by stress-testing it using a simulation program, or by developing the scheme program. The approach we are going to follow actually depends on the anticipated proposed scheme and it's complexity.

\hfill

The newly developed key distribution scheme is expected to provide faster search, insert, delete, and update operations because it's going to capitalize on B- tree algorithm, which has already been established and proven to be one of the best in term of performance.

\hfill

\section{APPLYING B-TREE CONCEPTS TO OPTIMIZE PEER TO PEER NETWORK}
B-Tree supports any P2P network to be balanced which makes the key distribution process more complex. However, it affords a fully guarantee to deliver the session key to all nodes at very short time comparing with the distribution key through unbalanced tree. In this project, we went through some assumptions to make the simulation goes perfectly. In following sections we will see some of these concepts or assumptions by some kind of details.

\hfill

\subsection{ RANKING NODES AND LOOKUP TABLE}
The first concept that we want to discuses is the value that the B-Tree will sort the nodes based on. The key value of our idea is calculation the nodes values based on the whole time that the node is being online in the P2P network. Based on this time, we give the nodes a sorting value which we named is as "Rank Value". The rank value is used as input value to the system and used by sorting algorithm to sort the B-Tree. Each node has to have a unique rank value, so we store the rank values in lookup table that the system uses to read them from. 

\hfill

One more important thing is, the online time is a cumulative value which means in case of leaving any node that time will be saved in the lookup table which will be stored in a server as database, we will discuss the idea of having this kind of server in section IV-C, and when the node joins the network the online time will be added to the old one. By doing that, we give the most trust ranking to the node who has the most online time value and so on.

\hfill

\subsection{JOINING/LEAVING NODES}
The operation of joining/leaving nodes has been discussed in section III. In addition, there are two differences points that we would mention about:
\subsubsection{REBALANCING THE TREE}
There are two situations that any node could be in the P2P tree: 

First situation, the node might be a leaf for a parent that has two leaves in the tree. So, with this case the system doesn't have to rebalance the tree because the leaf nodes don't make any changes on the distribution. However, if the parent only has one leaf, the tree needs to rebalanced again.

Second situation, the node might be a parent which means has children, so in this case when this node joins/leaves the tree, the system must rebalance the tree again.
\hfill

\subsubsection{REKEYING}
In our system, the KDC generates the session key every time the system needs it in any situation. For security purposes, the system must regenerate the session key and distribute it between nodes. So, the process time of informing the root node about the joining/leaving any node will take the same time of delivering the session key to the level of this node. 

\hfill

\subsection{CONTROLLING SERVER}
The idea of having controlling server is that the KDC will not be connected directly with the root node because the system might have a different roots at any duration of time. So, we need a server that control this process which is request the rekeying from KDC and does the process of distributing and rebalancing the P2P tree. Also, this server works partly as a database to store an updated copy from the lookup table of the nodes.
\hfill

\section{OUR SIMULATION}
The main scope of our project is balancing the tree based on the online time of each node. Although we calculated the ranking value based on the B-Tree algorithm, we changed the values of nodes to give the concept of having parents that have more value than the children whereas the original B-Tree concept is that the highest value be the right child and lowest value be the left child.

\hfill

In this section we will discuss the way that we have implemented our simulation by. The following sections discuss our work:

\hfill

\subsubsection {LOOKUP TABLE} We built the lookup table and give the simulator the ability to enter the number of nodes that the tree might have.This table has different values for each node as following:

- Online time: is generated randomly. 

- User ID: based on number of nodes.

- Ranking value: we calculate it by giving the middle value for the node who has the highest online time.

\hfill

\subsubsection {Calculating the levels of the tree} by using the following equation, we calculated the levels of the tree:

\begin{equation}
	O (log _{ d }{ \quad  { n\quad} })  
\end{equation}

Where:

- n: is the number of nodes.

- d: is number of children for the parent.

\hfill

\subsubsection {Implementation} by selecting the values of (n= 333, d=2) we got the following chart:
\hfill

\hfill

\includegraphics[width=90mm,height=60mm,scale=0.75]{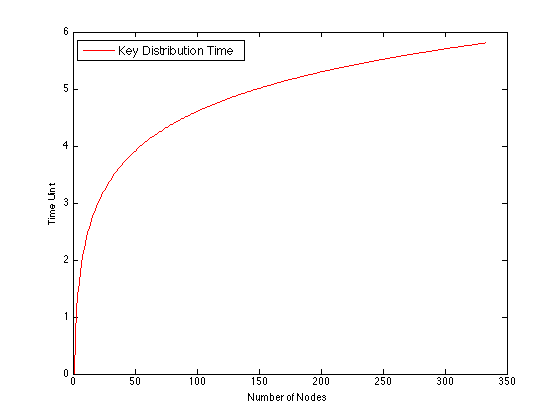}

Figure 1. Number of nodes that every level could cover.

\hfill

This chart illustrates the number of nodes for every unit time. Unit of time reflects the level number of the tree. By reading the chart we find, in level one (time one) only the two children could be covered and get the session key. In time four, there will be around 50 nodes or so could be covered and get the session key. So, for all nodes we need only six time units to cover all 333 nodes.
\hfill

\hfill

\section{RESULTS AND PERFORMANCE}
From applying the different numbers of nodes, we got different results that reflected the high performance of having B-Tree to distribute the session key in P2P network. In any unbalanced P2P network some nodes like leaves nodes could get the session key after log time because this node location might be in the based of the tree or so.

\hfill

The results give indications for the performance that we measured by doing the following. We calculated the performance of B-Tree which is calculated by equation number (2). Also, we added the performance of generating the session key by KDC and deliver it to the controlling server.

\hfill

There two values are represented by a time unit and give the whole performance of the system which is optimizing key distribution in P2P network using B-Tree.

\hfill

We have developed our simulation using Java programming language. As a result of that, we had to compromise having a high performance b-tree and settle with a data structure that applies the logic of b-tree on a list object. This because of the lack of pointers and memory address manipulation in Java. This, in fact, added some overhead to the proposed scheme. Such an overhead was successfully avoided in  \cite{Bianchi2010Stabilizing} by using an array of pointer to simulate their proposed distributed balanced tree, which is used to construct a peer-to-peer network optimized for selective dissemination of information.

\hfill

Another issue this simulation has raised is the effort the b-tree based proposed solution would take to rebalance itself. This task would exhaust the system available resources more with higher number of nodes. To mitigate this issue, we can configure and tune the system to maintain reasonable balance rate that doesn't affect the overall performance. This implies that the tree structure in the proposed scheme wouldn't be completely balance all the time. However, this shouldn't reach an unacceptable rate.

\hfill
\section{Conclusion}

Peer-to-Peer networks need to be more secure because of absence of centralization of controlling the communication between peers. This weakness caused by different effects, one of these is the time of delivering the session key to all nodes in a very close time to avoid the chance of having the opponent eavesdropping to the communication.

\hfill

This work concludes the high performance of using B-Tree to distribute the session key in Peer to Peer network by distributing the session key as less time as we can. For 333 nodes, we only need about 6 units time to deliver the key to all nodes. That makes the P2P more powerful and secure.

\hfill

The security service that we offer to the network is confidentiality by allowing all nodes using the session key in a time where the opponent can't get it because of the very short distributing time. Also, by making the nodes get the same session key before joining/leaving more than one node, which means make all nodes using the same session key for the same session and keep the communication to be synchronous.

\section*{Acknowledgment}

The authors would like to thank Professor. Wu, for his guidance throughout this research project. His constant remarks and constructive feedback were always valuable input for this works.


\bibliographystyle{IEEEtran}
\bibliography{BibDesk}

\end{document}